# The fallacy of sustainable Generative AI: limitations in EU environmental regulation of data centres and paths forward

Daria Onitiu*, Sandra Wachter** and Brent Mittelstadt***

## Abstract

In the age of Artificial Intelligence (AI), Large Language Models, Generative AI and larger frontier AI models, data centres create a significant environmental burden on electricity grids and fresh water resources. Requiring data centre operators and Big Tech under the recast Energy Efficiency Directive (recast EED) to quantify, report and disclose the facility-level energy and water impacts seems to be a step into the right direction towards more transparency and accountability. Yet when two recast EED approved benchmarks - the Power Usage Effectiveness (PUE) and Water Usage Effectiveness (WUE) - can be skewed to create a false sense on efficiency gains, current EU policy pushing for sustainable hyperscale data centre expansion appears misplaced. This paper argues that current PUE and WUE reporting frameworks illustrate what we term the "efficiency paradox," according to which positive scores require retrofitting larger AI data centres at the expense of energy supply and people's water access. Countering this efficiency paradox requires a new strategy for data centre operators and the EU Commission to demonstrate the ecological gains of optimising for efficiency through individual reporting and additional policy interventions. We make three policy proposals to show how this strategy can be formalised in practice: (i) measures to reveal PUE and WUE trade-offs relative to efficiency improvements, (ii) a comparative baseline for efficiency improvements securing actual ecological gains, and considering water stress, (iii) a monitoring framework of their diminishing returns and countereffects over time. Implementing these measures will ensure that the recast EED common rating scheme is fit-for-purpose, balancing sustainability with AI innovation for local communities.

## 1. Introduction

In the age of Artificial Intelligence (AI), Large Language Models (LLMs), Generative AI and larger frontier AI models, we have to look at the infrastructure that sustains their deployment: data centres. Data centres are physical "warehouses" housing stacks of servers, hardware and Graphic Processing Units (GPUs).[1] Big Tech companies such as Google, Microsoft, and Amazon are 'dominating' the infrastructure to support the exponential growth of AI, including hyperscale data centres.[2] These facilities, stretching '13 football fields',[3] carry a significant environmental burden with estimations on their electricity use exceeding Japan's total consumption[4] and with the significant water footprint being sometimes a tightly kept secret.[5] Set against a backdrop of vague EU policy seeking to balance sustainability and AI innovation in data centre infrastructure,[6] our paper analyses the facility-level reporting and transparency scheme under the recast Energy Efficiency Directive (recast EED)[7] and identifies interventions that help AI data centre operators demonstrate, and the EU Commission formalise the ecological gains of reducing their large energy footprint.

* Hasso-Plattner-Institute, Prof.-Dr.-Helmert-Str. 2-3, Potsdam, D-14482, Germany and Oxford Internet Institute, University of Oxford, Stephen A Schwarzman Centre for the Humanities, Oxford, OX2 6GG, UK. Email: Daria.Onitiu@hpi.de.
** Hasso-Plattner-Institute, Prof.-Dr.-Helmert-Str. 2-3, Potsdam, D-14482, Germany and Oxford Internet Institute, University of Oxford, Stephen A Schwarzman Centre for the Humanities, Oxford, OX2 6GG, UK.
*** Oxford Internet Institute, University of Oxford, Stephen A Schwarzman Centre for the Humanities, Oxford, OX2 6GG and Weizenbaum Institute, Hardenbergstraße 32, Berlin, 10623, Germany.
[1] Jessica Commins and Kristina Irion, 'Towards Planet Proof Computing: Law and Policy of Data Centre Sustainability in the European Union' (2025) 2025 Technology and Regulation 1, 6.
[2] Agnieszka Widuto, 'AI and the Energy Sector' [2025] (PE 775.859) 3.
[3] 'This Is the State of Play in the Global Data Centre Gold Rush' (*World Economic Forum*, 22 April 2025) <https://www.weforum.org/stories/2025/04/data-centre-gold-rush-ai/> accessed 1 November 2025.

This paper makes two contributions. First, it notes that recast EED endorsed metrics − the Power Usage Effectiveness (PUE) and Water Usage Effectiveness (WUE) metrics − are not a direct measure of efficiency while also incentivising data centre operators to upscale larger and more power-hungry facilities. The PUE and WUE are benchmarks for operators to measure the electricity and operational water consumption of the facility and its IT infrastructure, with lower scores indicating higher efficiency.[8] With an increasing IT load and infrastructural demand to cool denser AI servers and hotter racks, hyperscale operators need to utilise "technical optimisation measures" directed to improve the energy efficiency of its IT equipment and cooling infrastructure.[9] While technical optimisation measures may bring down PUE/WUE scores temporarily, the reading of these scores can mask important context due to behavioural and economic changes in the demand for AI services (e.g., rebound effects).[10] Further, technical optimisation of IT equipment alone will not maintain these small, marginal gains in PUE and WUE scoring and requires operators to apply aggressive optimisation in the cooling infrastructure. In practice, this means significant retrofitting of the facility and comes with a significant cost for local energy grids and peoples' access to fresh and potable water. These considerations point to a broader problem, which we describe as the "efficiency paradox on data centre efficiency". Data entre operators can boast low PUE and WUE values, with efficiency improvements being only marginal in relation to AI models' increasing computational demands, ignoring second-order effects. Ultimately, this is driving further data centre expansion at the expense of local communities' energy and water access.

---

4 Davide D'Ambrosio, Hugh Hopewell, Vincent Jacamon, Alex Martinos, Nichlas Salmon and Brent Wanner, 'Energy and AI' (International Energy Agency, 2025) 33 < https://www.iea.org/reports/energy-and-ai> accessed 2 October 2025.

5 Luke Barrat and Rosa Furneaux, 'Amazon strategised about keeping its datacentres' full water use secret, leaked document shows' (*The Guardian*, 25 October 2025) < https://www.theguardian.com/technology/2025/oct/25/amazon-datacentres-water-use-disclosure> 22 April 2026; Nuoa Lei, Jun Lu, Araman Shebabi and Eric Masanet, 'The Water Use of Data Center Workloads: A Review and Assessment of Key Determinants' [2025] 219 Resources, Conservation and Recycling 1-2.

6 Commins and Irion (n1) 12-13, 23-25; Thomas Le Goff, 'The Unsuitability of Existing Regulations to Reach Sustainable AI' (Knowledge to Action: Strategic Dialogue towards COP30, Rio de Janeiro, Brazil, 2025) 4; Philipp Hacker, 'Sustainable AI Regulation' (2024) 61 (2) Common Market Law Review 345-386; Philipp Hacker, Kai Ebert, Nicolas Alder, Vlad C Coroamă, Ralf Hebrich, 'Policy Brief on Powering Europe's Digital Transformation: A Roadmap to Greener Data Centres' ACM Europe Technology Policy Committee, 2 June 2025) <https://www.acm.org/binaries/content/assets/public-policy/europe-tpc/acm_data_center_final.pdf> accessed 23 September 2025; Hannah Ruschemeier, 'Climate Governance and AI Regulation: Bringing Together Two Sides of the Same Coin?' (2026) 24 (1) Zeitschrift für Europäisches Umwelt- und Planungsrecht 11, 17; Philip Hacker, Nicolas Alder, Kai Ebert and Ralf Hebrich, 'Policy Brief on The EU AI Act for True Environmental Accountability' (ACM Europe Technology Policy Committee, 2 June 2025) 5 <https://www.acm.org/binaries/content/assets/public-policy/europe-tpc/acm_climate_disclosure_final.pdf> accessed 21 June 2025; Kai Ebert, Boris Gamazaychikov, Philipp Hacker and Sasha Luccioni, 'The Global Landscape of Environmental AI Regulation: From the Cost of Reasoning to a Right to Green AI' (ArXiv, 10 February 2026) 5 <https://doi.org/10.48550/arXiv.2603.00068> accessed 26 March 2026; cf Kai Ebert, Nicolas Alder, Ralf Hebrich and Philip Hacker, 'AI, Climate, and Regulation: From Data Centers to the AI Act' [2026] 61 Computer Law & Security Review 1, 2-3, 6, 9; Nicolas Alder, Kai Ebert, Ralf Herbrich and Philip Hacker, 'AI, Climate, and Transparency: Operationalizing and Improving the AI Act' (2025) 14 (4) Journal of European Consumer and Market Law 166; Ajit Niranjan, 'US tech firms successfully lobbied EU to keep datacentre emissions secret' (*The Guardian*, 17 April 2026) < https://www.theguardian.com/technology/2026/apr/17/microsoft-us-tech-firms-lobbied-eu-secrecy-rules-datacentre-emissions> accessed 24 April 2026.

7 Directive (EU) 2023/1791 of the European Parliament and of the Council of 13 September 2023 on energy efficiency and amending Regulation (EU) 2023/955 (recast) [2023] OJ L 231/1 (hereafter *recast EED*).

8 BS EN 50600-4-2:2016+A1:2019 Information Technology- Data Centre Facilities and Infrastructures, Part 4-2: Power Usage Effectiveness [2019] s 5.1; BS EN 50600-4-9:2022 Information Technology- Data Centre Facilities and Infrastructures Part 4-9: Water Usage Effectiveness [2023] s 5.

9 Thomas Le Goff, Oliver Inderwildi, Friðrik Már Baldursson, Nils-Henrik M von der Fehr, 'From Gridlock to Grid Asset: Data Centres for Digital Sovereignty, Energy Resilience, and Competitiveness' (Centre on Regulation in Europe (Cerre), 2025) 22 < https://cerre.eu/publications/from-gridlock-to-grid-asset-data-centres-for-digital-sovereignty-energy-resilience-and-competitiveness/> accessed 4 October 2025; D'Ambrosio and others (n 4) 47.

Second, the paper develops three concrete measures that address this mismatch between reported efficiency and ecological gains under the recast EED.[11] These entail (i) measures to reveal PUE and WUE trade-offs relative to efficiency improvements, (ii) a comparative baseline for efficiency improvements securing actual ecological gains, and considering water stress, (iii) a monitoring framework of their diminishing returns and countereffects over time. Our findings underscore a more environmentally consistent understanding of data centre efficiency, not merely as market-aligned performance indicators, but as a yardstick contextualising PUE and WUE scores against growth in AI demand and data centre capacity.

## 2. Why we need a new approach to data centre efficiency

With surging demand for resource-intensive AI services and products, a reliable strategy to asses and plan data centre sustainability is needed. The following Sections show why scaling to larger and increasingly power-hungry facilities is a necessary precondition for Big Tech to showcase environmental gains, focusing on the PUE and WUE metrics. These are methodologies for data centre operators to measure the facilities' electricity consumption, as well as the infrastructure's operational water consumption.[12]

One would reasonably expect that an increased AI training and use (i.e., inference) capacity would put pressure and increase both PUE and WUE scores. Yet, Big Tech companies are communicating *low* PUE and WUE scores for their large facilities.[13] The reason is the interplay between two effects illustrated in **Figure 1**: (i) the *aggregate effects* of technical optimisations yielding diminishing returns of data centre efficiency, and (ii) the marginal gains in PUE and WUE scoring and reporting of *individual* data centre efficiency after technical optimisations have been put in place. The interplay between these two factors underscores an efficiency paradox where further retrofitting of larger facilities is required to maintain smaller, marginal PUE and WUE gains.

---

[10] Alexandra Sasha Luccioni, Emma Strubell and Kate Crawford, 'From Efficiency Gains to Rebound Effects: The Problem of Jevons' Paradox in AI's Polarized Environmental Debate' (Proceedings of the 2025 ACM Conference on Fairness, Accountability, and Transparency (Association for Computing Machinery, Athens, Greece, 2025) <https://doi.org/10.1145/3715275.3732007> accessed 18 April 2026.
[11] *recast EED* (n 7) art 12.
[12] BS EN 50600-4-2:2016+A1:2019 (n 8); BS EN 50600-4-9:2022 (n 8).
[13] 'Amazon Sustainability Report' (AWS, 2024) 3-4 <https://sustainability.aboutamazon.com/2024-amazon-sustainability-report-aws-summary.pdf> accessed 16 January 2026; Google, 'Power Usage Effectiveness: Growing the Internet While Reducing Energy Consumption' (*Google Data Centers*) <https://datacenters.google/efficiency> accessed 14 November 2025; Meta, 'Environmental Data Index. 2025 Sustainability Report (Reporting on Fiscal Year 2024)' (2025) 3, 8 <https://sustainability.atmeta.com/wp-content/uploads/2025/10/Meta_2025-Environmental-Data-Index.pdf> accessed 16 January 2026.See also, Steve Solomon, 'Sustainable by Design: Next-Generation Datacenters Consume Zero Water for Cooling' (*The Microsoft Cloud Blog*, 9 December 2024) <https://www.microsoft.com/en-us/microsoft-cloud/blog/2024/12/09/sustainable-by-design-next-generation-datacenters-consume-zero-water-for-cooling/> accessed 4 September 2025.

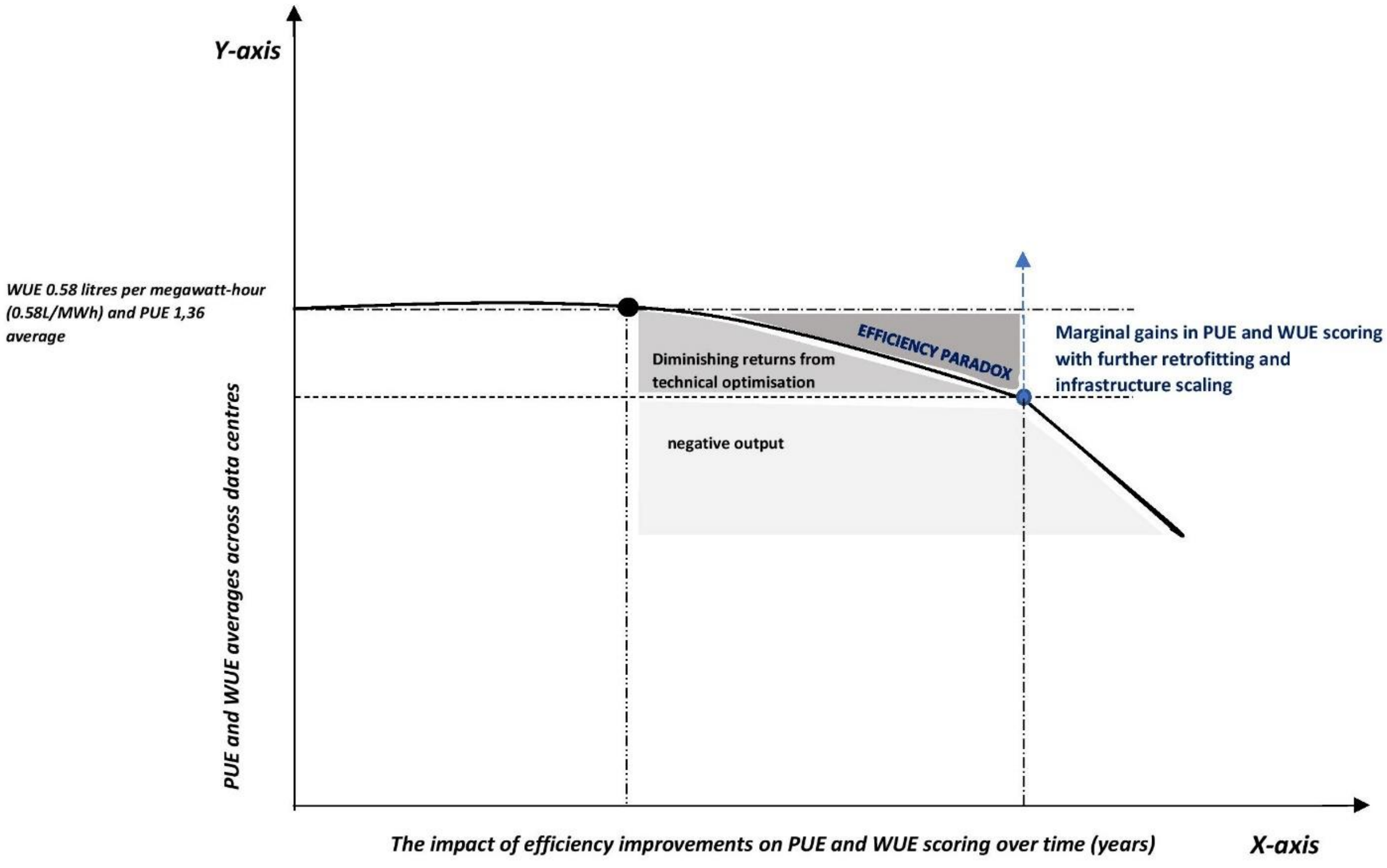

**Figure 1**. This figure illustrates the interplay between technical optimisation measures to achieve facility-level energy efficiency and how their aggregated impact and reported Power Usage Effectiveness (PUE) and Water Usage Effectiveness (WUE) scores can provide a skewed understanding of environmental sustainability. The Y-axis provides two illustrative PUE and WUE averages taken from technical reports on the recast Energy Efficiency Directive. The X-axis underscores how technical optimisations translate to diminishing returns and marginal gains in efficiency over a significant period of time. Specifically, when technical optimisations, such as processor-level improvements in IT equipment, only translate small gains in improving facility-level energy efficiency, then these optimisations yield diminishing returns. The blue arrow then highlights that to maintain these small, marginal gains in PUE and WUE reporting, requires further scaling in the cooling infrastructure. The alternative would be a negative output over time when increases in AI workload and demand are not matched with aggressive technical optimisation. Hence, maintaining PUE and WUE marginal gains, requiring upscaling to larger facilities signifies an efficiency paradox where it pays off to roll-out more power-hungry facilities.

To define the interplay between diminishing returns and marginal gains, let us use an illustrative example. Imagine a facility that wants to shift to fully handling large AI workloads. Assume that the current PUE and WUE average across mid- and large-scale data centres is already relatively low, entailing a WUE of 0.58 litres per megawatt-hour (L/MWh) and a PUE of 1.36, as noted on the *y-axis* in Figure 1.[14] Nevertheless, to further improve its facility-level energy efficiency (i.e., moving

[14] It is difficult to outline current industry-average PUE and WUE values due to lack of comprehensive PUE and WUE data insights. We use initial findings from the technical report on the recast European (EU) Energy Efficiency Directive to at least approximate prevailing EU trends entailing a WUE of 0.58 L/kwh and PUE of 1.36. For data centres with a capacity of above 10MW, the current PUE average for instance can be indeed lower (i.e., PUE 1.21), Dennis Heatubun, Ionut Dragan, Marco Bolchi, Patricia Lopes Bautista, Simon Hinterholzer, Ralph Hintemann, Nicolas Marx, Patrik Furda, 'Assessment of the Energy Performance and Sustainability of Data Centres in EU: First Technical Report' (EU Commission, 2025) 81,39 < https://op.europa.eu/da/publication-detail/-/publication/83be4c3e-5c79-11f0-a9d0-01aa75ed71a1/language-en> accessed 4 September 2025. See also, PUE targets in national jurisdictions (i.e., China, Singapore and

upwards the *y-axis*), the individual data centre operator may employ a range of what we term "technical optimisation" measures. For example, batching is a technical optimisation method that allows for more efficient utilisation of Graphic Processing Units (GPUs) by grouping tasks together, rather than processing them individually.[15] While this method can lead to more efficient utilisation and energy use of GPUs, these energy gains can yield diminishing returns including smaller gains in efficiency when 'increasing the batch size'.[16] Additionally, there are other, second-order effects of technical optimisations' diminishing returns.[17] Using batching as an example, if this optimisation method were applied to the 'scaling of AI' for more end-users, the resulting increase in demand would again raise overall energy consumption.[18]

Further, our data centre needs to further optimise the IT equipment and infrastructural energy use of its facility to avoid its PUE and WUE plateauing or even attaining a negative output. While there is no openly accessible data that underscores overall data centre inefficiency, a negative output *could* occur If the data centre does not invest in *any further* technical optimisation on an infrastructural level for instance while significantly increasing the AI workload(s). These considerations would have serious implications: the data centre could be classified as a 'worst-performing facility',[19] unable to significantly progress in lowering PUE and WUE scores.

A next step for the data centre would therefore be to maintain the marginal, small gains in the PUE and WUE scores through further retrofitting of its infrastructure. In simple terms, the optimisation of the IT equipment, such as batching, needs to be matched with aggressive optimisation of the facilities' cooling infrastructure. Conversely, simply continuing optimising the IT equipment will not improve PUE and WUE scores because these are ratio metrics: the data centre's annual electricity and water consumption is relative to the energy consumed by the IT equipment (e.g., servers).[20] The upshot of this is that retrofitting a facility this way, for example, by matching the larger processing batches with more GPUs and cooling hotter racks, enables the data centre to maintain its current low PUE and WUE averages while still reporting improved PUE and WUE scores (with an eye towards future efficiency targets).

Although the interplay between individual, ratio-effects in PUE and WUE scoring (on the *y-axis*) and systemic effects on an aggregate level (on the *x-axis*) entails complex measurements,[21]

---

Germany), that are close (i.e., PUE 1.2-1.3) to the stipulated average, Fiona Brocklehurst, 'Policy Development on Energy Efficiency of Data Centres' (Electronic Devices & Networks Annex (EDNA), 2024) 13-14 < https://www.iea-4e.org/wp-content/uploads/2024/02/Policy-development-on-energy-efficiency-of-data-centres-draft-final-report-v1.05.pdf> accessed 24 October 2025.

[15] D'Ambrosio and others (n 4) 47.

[16] D'Ambrosio and others (n 4) 47; Nidhal Jegham, Marwan Adbdelatti, Chan Young Koh, Lassad Elmoubarki, Abdeltawab Hendawi, 'How Hungry Is AI? Benchmarking Energy, Water, and Carbon Footprint of LLM Inference' (*ArXiv*, 24 November 2025) 12 <https://doi.org/10.48550/arXiv.2505.09598> accessed 23 January 2026.

[17] Luccioni, Strubell and Crawford (n 10); Commins and Irion (n 1) 24-25.

[18] Luccioni, Strubell and Crawford (n 10) 5.

[19] Simon Hinterholzer, Ralph Hintemann, Severin Beucker, Nicolas Marx, Patrik Furda, Dennis Heatubun, Ionut Dragan, Marco Bolchi, Patricia Lopes Bautista, 'Assessment of next Steps to Promote the Energy Performance and Sustainability of Data Centres in EU, Including the Establishment of an EU-Wide Rating Scheme: Second Technical Report' (EU Commission, 2025) 67 <https://op.europa.eu/en/publication-detail/-/publication/29aa14ed-a4ba-11f0-a7c5-01aa75ed71a1/language-en> accessed 12 November 2025.

[20] BS EN 50600-4-2:2016+A1:2019 (n 8) D.2; BS EN 50600-4-9:2022 (n 8) s 3.1.2; BS ISO/IEC 30134-4:2017+A1: 2025 Information technology – Data centres- Key performance indicators Part 4: IT Equipment Energy Efficiency for servers (ITEEsv) s 3.1.1-3.1.2.

[21] Peter Gailhofer, Anke Herold, Jan Peter Schemmel, Cara-Sophie Scherf, Cristina Urrutia, Andreas R. Köhler and Sibylle Braungardt, 'The Role of Artificial Intelligence in the European Green Deal' (Study requested by the AIDA Committee, 2021) (PE 662.906) 10 <https://www.europarl.europa.eu/RegData/etudes/STUD/2021/662906/IPOL_STU(2021)662906_EN.pdf> accessed 25 May 2025.

underscoring this relationship is nonetheless important. While PUE and WUE are facility-level metrics,[22] second-order effects, such as behavioural and economic changes including rebounds are relevant especially for policy makers, because these can effectively 'offset efficiency improvements'.[23] In this regard, some data already indicates diminishing efficiency gains in areas such as PUE scoring or AI-accelerated computing;[24] however, we have not set a specific date when broader diminishing returns in (technical) efficiency improvements would occur in Figure 1 (including the *x-axis*). Establishing such a timeline would require a top-down or bottom-up analyses[25] of AI accelerated computing, AI training and deployment trends,[26] being beyond the scope of this paper. Instead, the visualisation shows that once efficiency improvements plateau and decrease, any marginal gains need to be outmatched with further investments in IT equipment and cooling that are typically designed to sustain larger AI capacities. This reasoning is in line with current data that future efficiency gains in AI require further optimisations to sustain efficiency gains, such as adopting advanced cooling solutions.[27]

Retrofitting data centres into larger facilities capable of handling greater AI workloads comes with a significant environmental price tag for local communities. The global electricity demand of data centres is currently estimated to reach 945 terawatt-hours (TWh) by 2030, a comparable share that exceeds the consumption of entire Japan.[28] The increased power capacity and demand further exacerbates pressure on local energy grids and can lead to unstable power supply for residents.[29] In Aragón, Spain, the roll-out of 24 large data centres in recent years meant that the 'regional government had to put [in] some grants for the farmers because there is such a drought that the crops are dying'.[30]

However, data centre operators do not need to account for these environmental impacts, the diminishing returns and second-order effects of technical optimisations, or the fact that PUE and WUE reporting will deliver only marginal efficiency gains as the facility expands. The core of the problem is an efficiency paradox where in practice data centre operators can expand their facilities endlessly but maintain (or even reduce) their average PUE/WUE across their facilities. This is the sense in which the reading and reporting of PUE/WUE scores fail to capture the true environmental impact of data centres, particularly on localised regions and communities: they can accurately capture efficiency averages across computational equipment and facilities, but obfuscate or ignore its impact on local and global energy and water resources.

---

[22] Amin Iszadeh, Davide Ziviani and David E Claridge, 'Global trends, performance metrics, and energy reduction measures in datacom facilities' [2023] 174 Renewable and Sustainable Energy Reviews 1, 9.
[23] Ruschemeier (n 6) 20.
[24] George Kamiya and Vlad C Coroamă, 'Data Centre Energy Use: Critical Review of Models and Results' (Efficient, Demand Flexible Networked Appliances (EDNA), 2025) 30 <https://www.iea-4e.org/wp-content/uploads/2025/05/Data-Centre-Energy-Use-Critical-Review-of-Models-and-Results.pdf> accessed 16 January 2026; Vlad Coroamă and Simon Hinterholzer, 'Energy Efficiency of Servers' (Electronic Devices & Networks Annex (EDNA), 2025) 34 <https://www.iea-4e.org/wp-content/uploads/2025/05/EDNA-EE-of-servers-FINAL.pdf> accessed 5 November 2025; Douglas Donnellan, Andy Lawrence, Daniel Bizo, Peter Judge, Jon O'Brien, Jacqueline Davis, Max Smolaks, Jabari Williams- George, Rose Weinschenk 'Uptime Institute Global Data Center Survey 2024' (Uptime Institute 2024) 7 < https://datacenter.uptimeinstitute.com/rs/711-RIA-145/images/2024.GlobalDataCenterSurvey.Report.pdf> accessed 18 April 2026; D'Ambrosio and others (n 4) 47; Le Goff and others (n 9) 22.
[25] Kamiya and Coroamă (n 24) table 3.4.
[26] Coroamă and Hinterholzer n (24) 24.
[27] United Nations Environment Programme (UNEP) United for Efficiency (U4E) initiative, 'Sustainable Procurement Guidelines for Data Centres and Servers' (United Nations Environment Programme (UNEP) United for Efficiency (U4E) initiative, 2025) 19 < https://wedocs.unep.org/handle/20.500.11822/47717> accessed 26 March 2026.
[28] D'Ambrosio and others (n 4) 49.
[29] Widuto (n 2) 3, 5.
[30] Mathieu Pollet, 'The EU Wants to Build Thousands More Data Centers. Where Will They Go?' (*POLITICO*, 23 March 2026) <https://www.politico.eu/article/eu-data-center-where-will-they-go/> accessed 29 March 2026.

## 2.1. Measuring data centre electricity and water footprint

Before explaining why data centre efficiency is skewed towards environmentally inconsistent behaviour by major cloud and AI providers, we first outline how the PUE and WUE are measured. The PUE and WUE are two important metrics for data centre operators to quantify the electricity and water footprint of data centres.[31] These ratio-metrics capture its total data centre energy consumption, entailing its infrastructure, such as cooling system and its IT equipment (e.g., servers) and is divided by the IT equipment to store, manage and process data.[32]

The calculation of the PUE score usually follows this formula:

$$\text{Power Usage Effectiveness (PUE)} = \frac{\text{Annual total Annual total data centre energy consumption in kilowatt per hour (kWh)}}{\text{Annual IT equipment energy consumption in kilowatt per hour (kWh)}}$$

The WUE score is usually calculated in the following way:

$$\text{Water Usage Effectiveness (WUE)} = \frac{\text{Annual water usage of the data centre in m3}}{\text{Annual IT equipment energy consumption in MWh}}$$

The following discussion will set the scene for the efficiency paradox through a brief outline what PUE and WUE intend to measure, how these metrics differ from each other and their limitations. In this regard, both are facility-level metrics only, with the PUE pertaining to facilities' relative infrastructural efficiency and the WUE not capturing total water use and consumption.[33] Despite these metric's limitations, these metrics a remain 'convenient' to report power overheads, as well as drive the 'operational site's water use'.[34] Moreover, regulations still 'assign a central role' to these values, especially the PUE.[35] Hence, the following discussion intends to appropriately carve out PUE/WUE's limited operational scope, which in turn, sets the tone of why their "marketing" can

[31] BS EN 50600-4-2:2016+A1:2019 (n 8) s 4, 3.1.3; BS EN 50600-4-9:2022 (n 8) s5, 3.1.3.

[32] BS EN 50600-4-2:2016+A1:2019 (n 8) s 5; BS EN 50600-4-9:2022 (n 8) s 5.

[33] Lauren E Bridges, 'Environmental Impacts of AI' in Rayvon Fouché (ed), *Oxford Research Encyclopedia of Science, Technology and Society* (Oxford Academic 2026) 8.

[34] Jumie Yuventi and Roshan Mehdizadeh, 'A critical analysis of Power Usage Effectiveness and its use in communicating data center energy consumption' [2013] 64 Energy and Buildings 90, 93; Michael Patterson, 'Water Usage Effectiveness (WUE™): A Green Grid Data Center Sustainability Metric' (The Green Grid, 2011) 7 <https://www.thegreengrid.org/system/files/store/WUE_v1.pdf> accessed 2 October 2026.

[35] Vlad C. Coroamă and Oana Dumbravă, 'Circular Economy Synergies and Trade-offs in Data Centres' (CEST-DC, 2026) 20, 38 < https://arxiv.org/abs/2608.04571> accessed 2 October 2026; See also, Bridges (n 33) 8.

obfuscate complex dependencies and broader governance challenges surrounding AI's higher energy demand**.**

Finally, an important caveat is warranted at this stage: external observers such as researchers cannot deduce *how much* of the processing in a data centre is specifically devoted to AI workloads.[36] Further, data centre operators themselves are often unable or unwilling to disclose the composition of their IT workload.[37] Nevertheless, we can focus on the elements composing the PUE and WUE scores to interpret how (larger) AI models *could* put increasing pressure on PUE and WUE values compared to traditional data processing tasks due to their unique specifications for IT equipment, energy intensive training and inference, and resulting high cooling demands. Exploring these relationships within PUE and WUE metrics provides a helpful starting point to understand how data centre operators lower and direct energy efficiencies to meet rapidly scaling demand on data centres. We will then be able to precisely define the gaps that exist when marginal gains in PUE and WUE scores are reported by data centre operators.

### 2.1.1. An outlook how Power Usage Effectiveness (PUE) and Water Usage Effectiveness (WUE) are measured

Understanding the scale and intensity of data centres' electricity and water footprints depends on how these impacts are measured. We will use a hypothetical illustration of how data centre operators would quantify the efficient use of energy and water using PUE and WUE.

Imagine a data centre which reports that for every 1 kWh consumed by its IT equipment (e.g., servers, data storage, network), 0,12 kWh are used for its overhead, such as lighting and cooling. To minimise the electricity consumption, the data centre operator would want to limit the infrastructure electricity consumption for cooling or lighting, bringing its closer to the IT equipment electricity consumption. The data centre operator might want to consider the facility's location and how more electricity consumption for cooling might have to work harder during summer seasons.[38] This effectively is an illustration of the PUE metric with a hypothetical score of 1.12. The score entails the ratio of the total facility's electricity consumption to the electricity consumption of the IT equipment.[39] A ratio of 1 would illustrate a theoretical ideal where the total facility's electricity consumption would entail the IT equipment, with no overhead electricity.[40] Because of seasonal and geographical variability, data centre operators would need to consider the data centre's operational efficiency over a period of time (e.g., 12 months).[41] Moreover, some counties including China have set different PUE targets for data centres located in warmer and cooler regions to reflect this variability.[42]

Now assume that powering its cooling infrastructure requires not only electricity but also water. Specifically, the hyperscale data centre employs a series of cooling technologies consuming mostly fresh and potable water to avoid overheating of its IT equipment, as well as the clogging of the cooling system.[43] Depending on the cooling technology, the facility's location and climate, the data centre might have to use more water.[44] By way of illustration, a data centre operating in a hot or dry environment, might adopt a cooling strategy entailing evaporative cooling (towers) to lower the outside air temperature.[45] This process of water evaporation, requiring a significant amount of water permanently removed from its source, can cause a high WUE score.[46] Other cooling strategies utilise outside air to cool down the IT equipment using a heat exchanger when the air temperature is low enough.[47] This cooling strategy might entail less or no water at all, but requires more energy due to the heat exchanger.[48] While the WUE score provides important insights on how cooling can entail varying amounts of water, it must be read in conjunction with the trade-offs of specific cooling systems for electricity consumption and the PUE value.[49]

---

[36] Sasha Luccioni, Boris Gamazaychikov, Theo Alves da Costa and Emma Strubell, 'Misinformation by Omission: The Need for More Environmental Transparency in AI' (*ArXiv*, 18 June 2025) 1 <https://doi.org/10.48550/arXiv.2506.15572> accessed 13 January 2026.

[37] Le Goff and others (n 9) 22.

[38] Nuoa Lei and Eric Masanet, 'Statistical Analysis for Predicting Location-Specific Data Center PUE and Its Improvement Potential' (2020) 201 (15) Energy 1, 6.

[39] BS EN 50600-4-2:2016+A1:2019 (n 8) s 3.1.2.

[40] Gemma A Brady, Nikil Kapur, Jonathan L Summers, Harvey M Thompson, 'A Case Study and Critical Assessment in Calculating Power Usage Effectiveness for a Data Centre' [2013] 76 Energy Conversion and Management 155, 156.

[41] BS EN 50600-4-2:2016+A1:2019 (n 8) s 5.1.

[42] Katerina Simou and Corina Bolintineanu, 'Data Centre Flexibility in Germany and China Status Quo and Best Practices' (Deutsche Energie-Agentur, 2022) 8 <https://www.dena.de/fileadmin/dena/Dokumente/Projektportrait/Projektarchiv/Entrans/Data_Centre_Flexibility_in_Germany_and_China_EN.pdf> accessed 13 November 2025.

In addition to the above considerations, there are important limitations in PUE and WUE scoring and reporting. By way of illustration, imagine an operator wants to optimise its IT load by consolidating its number of physical servers into virtual machines (also called virtualisation).[50] In this situation, the facility's PUE score would not accurately reflect the overall IT load reduction, as it does not provide 'insight into the productivity of the IT equipment'.[51] Indeed, one could think about complementing the PUE metric with additional methodologies, such as measuring the useful work per unit of energy using floating-point operations (FLOPs).[52] While such changes could lead to more comprehensive reporting, it is unlikely that Industry will abandon the PUE "target" altogether.[53]

---

[43] Shaolei Ren, 'How Much Water Does AI Consume? The Public Deserves to Know' (*OECD.AI Policy Observatory*, 30 November 2023) <https://oecd.ai/en/wonk/how-much-water-does-ai-consume> accessed 8 June 2025.

[44] Jens Gröger, Felix Behrens, Peter Galihofer, Inga Hilbert, 'Environmental Impacts of Artificial Intelligence' (Greenpeace, 2025) 24 < https://www.oeko.de/en/publications/environmental-impacts-of-artificial-intelligence/> accessed 29 August 2025.

[45] ibid 22.

[46] Andrew Higgins, 'What Is Water Usage Effectiveness (WUE) in Data Centers?' (*Interconnections - The Equinix Blog*, 13 November 2024) <https://blog.equinix.com/blog/2024/11/13/what-is-water-usage-effectiveness-wue-in-data-centers/> accessed 1 November 2025.

[47] Gröger and others (n 44) 22.

[48] Arman Shehabi, Alex Newkirk, Sarah J Smith, Alex Hubbard, Nuoa Lei, Md Abu Bakar Siddik, Billie Holececk, Jonathan Koomey, Eric Masanet and Sale Sartor, '2024 United States Data Center Energy Usage Report' (Berkley Lab Energy Analysis and Environmental Impacts Division, Lawrence Berkeley National Laboratory, 2024) 45 < https://escholarship.org/uc/item/32d6m0d1 > accessed 3 November 2025.

[49] Heatubun and others (n 14) 18; see also, Husam Alissa, Teresa Nick, Ashish Raniwala, Alberto Arribas Herranz, Kali Frost, Ioannis Manousakis, Kari Lio, Brijesh Warrier, Vaidehi Oruganti, TJ DiCaprio, Kathryn Oseen-Senda, Bharat Ramakrishnan, Naval Gupta, Ricardo Bianchini, Jim Kleewein, Christian Belady, Marcus Fontoura, Julie Sinistore, Mukunth Natarajan, 'Using Life Cycle Assessment to Drive Innovation for Sustainable Cool Clouds' (2025) 641 Nature 331, 332.

[50] International Telecommunication Union, 'SERIES L: Environment and ICTs, Climate Change, e-Waste, Energy Efficiency; Construction, Installation and Protection of Cables and Other Elements of Outside Plant Energy Efficiency, Smart Energy and Green Data Centres: Energy efficiency in micro data centres for edge computing' (International Telecommunication Union, 2024) Recommendation ITU-T L-1307 (03/2024) 7 < https://www.itu.int/itu-t/recommendations/rec.aspx?rec=15767> accessed 17 January 2026.

[51] BS EN 50600-4-2:2016+A1:2019 (n 8) s D.2; Alder and others 'AI, Climate, and Transparency: Operationalizing and Improving the AI Act' (n 6) 167-168; Nathaniel Horner and Inês Azevedo, 'Power Usage Effectiveness in Data Centers: Overloaded and Underachieving' (2016) 29 (4) The Electricity Journal 61.

[52] Le Goff and others (n 9) 22; Alexandre Lacoste, Alexandra Luccioni, Victor Schmidt and Thomas Dandres, 'Quantifying the Carbon Emissions of Machine Learning' (*ArXiv*, 4 November 2019) 4 <https://doi.org/10.48550/arXiv.1910.09700> accessed 6 June 2025; Roy Schwartz, Jesse Dodge, Noah A Smith, Oren Etzioni, 'Green AI' (2020) 63 (12) Communications of the ACM 54, 59–60.

[53] E.g., Viegand Maagoe and COWI, 'Reporting Requirements on the Energy Performance and Sustainability of Data Centres for the Energy Efficiency Directive Task B Report: Labelling and Minimum Performance Standards Schemes for Data Centres' (European Commission- Directorate-General for Energy, 2023) Section 4

WUE scoring faces similar limitations when data centres use water from many sources, for example reclaimed greywater, seawater, and potable water.[54] While WUE scoring can help operators to distinguish between different water categories, the standard does not mandate detailed links to specific water sources, and their implications for water stress in local communities.[55] Again, data centre operators could draw from additional metrics and region-specific data to complement WUE values.[56] Nevertheless, WUE requires additional context to understand the impact of a facility's water footprint on local environmental conditions.[57]

We are now able to define a key limitation of PUE and WUE scoring: they cannot, as currently formulated, provide comprehensive insights into the effectiveness of technical optimisation measures, such as virtualisation or cooling efficiency. While both metrics remain convenient and simple indicators that could be combined with other sustainability metrics, their reporting and regulatory reliance can create a loophole that enables endless data centre expansion without a similar reportable increase in environmental impact. Nevertheless, before elaborating these limitations and required solutions, we need to first discuss how increased training and use of larger AI models could exacerbate a facility's electricity and water consumption.

### 2.1.2. Trends in Large Generative AI workloads

Now imagine that our illustrative data centre processes a range of digital services but wants to completely shift to processing Large Generative AI Models. This would require significant retrofitting of their existing facility or even building a new one, becoming bigger in 'size and capacity'.[58] With the number of these hyperscale data centres continuously rising, complex and larger AI models put pressure on two items in the PUE and WUE scores, increasing IT equipment electricity consumption and operational water consumptions for cooling.

Meeting the demand for Large Generative AI training and deployment requires an increasing IT equipment load in our data centre. Additional and denser server racks and specialised hardware are needed to train and run these models. OpenAI, for example, has 'scaled up their clusters to 7,500 Graphic Processor Units [GPUs] to train LLMs, such as GPT-3'.[59] These GPU clusters are grouped in racks to enable parallel processing, shifting power capacity per rack, and increasing the

---

<https://energy.ec.europa.eu/topics/energy-efficiency/energy-efficiency-targets-directive-and-rules/energy-efficiency-directive_en> accessed 1 September 2025; Cf International Telecommunication Union (n 50) 5- 6.

[54] Shaolei Ren and Amy Luers, 'There Are Many Ways to Minimise Water Use Related to AI Operations, but They May Not Be What You Think' (*OECD.AI Policy Observatory*, 7 October 2025) <https://oecd.ai/en/wonk/ways-to-minimise-water-use-related-to-ai-operations-not-what-you-think> accessed 17 January 2026.

[55] Heatubun and others (n 14) 18, 40.

[56] Michael Patterson, 'Water Usage Effectiveness (WUE TM): A Green Grid Data Center Sustainability Metric' (Green Grid, 2021) 6, 9 <https://airatwork.com/wp-content/uploads/The-Green-Grid-White-Paper-35-WUE-Usage-Guidelines.pdf> accessed 2 September 2025; 'Climate Neutral Data Center Pact Initiative for Climate Neutral Data Centers in Europe' (2025) 2 < www.climateneutraldatacentre.net/wp-content/uploads/2025/06/CNDCP-Initiative-2025.pdf> accessed 31 March 2026.

[57] Yanran Wu, Inez Hua and Yi Ding, 'Not All Water Consumption Is Equal: A Water Stress Weighted Metric for Sustainable Computing' (2025) 5 (2) SIGENERGY Energy Informatics Review 84, 86. On how data centre's water footprint is only captured 'indirectly' in the EU Water Framework Directive, Cf Hacker (n 6) 355.

[58] Laure De Roucy-Rochegonde and Adrien Buffard, 'AI, Data Centers and Energy Demand Reassessing and Exploring the Trends' (The French Institute of International Relations (Ifri), 2025) 13 <https://www.ifri.org/sites/default/files/2025-02/ifri_buffard-rochegonde_ai_data_centers_energy_2025_0.pdf> accessed 1 November 2025.

[59] Pratyush Patel, Esha Choukse, Chaojie Zhang, Íñigo Goiri, Brijesh Warrier, Nithish Mahalingam, Ricardo Bianchini, 'Characterizing Power Management Opportunities for LLMs in the Cloud' [2024] 3 Proceedings of the 29th ACM International Conference on Architectural Support for Programming Languages and Operating Systems 207; Emma Strubell, Ananya Ganesh and Andrew McCallum, 'Energy and Policy Considerations for Modern Deep Learning Research' (2020) 34 (9) Proceedings of the AAAI Conference on Artificial Intelligence 13693, 13695.

cooling requirements. [60] Accommodating a data centre for large generative AI training would raise the energy requirements in a way that could require as much electricity to power an entire city, such as San Francisco in the United States, for several days.[61]

The inference stage, including the process through which a large generative AI model responds to user prompts, is also particularly resource-intensive for our data centre's electricity and water footprint.[62] A single inference requires less energy than model training;[63] however, once the model is deployed, we might be looking at 'billions or even trillions' inferences.[64] While companies do not disclose the exact numbers,[65] estimations suggest that some LLMs "consume" 'a 500ml water bottle for every 10 to 50 queries'.[66]

Beyond expanding IT equipment, our data centre also needs to increase its cooling infrastructure to support large generative AI workloads and at the expense of a larger water footprint.[67] For instance, Mytton notes that a 'medium-sized data centre (15 megawatts (MW)) uses as much water as three average-sized hospitals'.[68] Conversely, a larger facility that facilitates AI training workloads can consume up to '5 million gallons per day' corresponding to the water use of a town inhabited by '10,000 to 50,000 people'.[69] To summarise, hosting large generative AI services in our hypothetical data centre would require significant scaling in IT equipment and infrastructure to sustain increased rates of electricity and water consumption.

### 2.1.3. Hyperscale data centres: from scale to (in)efficiency

Imagine now that the operator still receives a local permit to expand the facility, despite the above indications and estimates suggesting that the surge in data centres will exacerbate existing environmental concerns, including pressure on energy supply and people's access to fresh and potable water.[70] Entire housing projects in West London (United Kingdom), for example, have been delayed until 2035 to ensure sufficient capacity exists for data centres.[71] Moreover, hyperscale data centres can cause water shortages for residents, especially in water stressed regions where the demand for potable water exceeds supply.[72] However, hyperscale data centres also often boast *low* PUE and WUE values.[73] This begs a question: should our hypothetical data centre and similar projects be approved to scale if they can achieve low PUE and WUE values?

---

[60] Boris Gamazaychikov and Sasha Luccioni, 'Power, Heat, and Intelligence – AI Data Centres Explained' (*Hugging Face*, 5 November 2025) < https://huggingface.co/blog/sasha/ai-data-centers-explained> accessed 1 January 2026.

[61] This example refers to the training of Open AI's GPT- 4; James O'Donnell and Casey Crownhart, 'We Did the Math on AI's Energy Footprint. Here's the Story You Haven't Heard.' (*MIT Technology Review*, 20 May 2025) <https://www.technologyreview.com/2025/05/20/1116327/ai-energy-usage-climate-footprint-big-tech/> accessed 23 October 2025.

[62] Alex de Vries, 'The Growing Energy Footprint of Artificial Intelligence' (2023) 7 (10) Joule 2191, 2192; Jegham and others n (16) 2–3.

[63] Leona Verdadero, Ivana Drobnjak, Hristijan Bosilkovski, Zekun Wu, Emma Fischer, María Pérez- Ortiz, 'Smarter, Smaller, Stronger: Resource-Efficient Generative AI & the Future of Digital Transformation' (UNESCO, 2025) 12 <https://unesdoc.unesco.org/ark:/48223/pf0000394521> accessed 15 July 2025.

[64] Sasha Luccioni, Yacine Jernite and Emma Strubell, 'Power Hungry Processing: Watts Driving the Cost of AI Deployment?' (Proceedings of the 2024 ACM Conference on Fairness, Accountability, and Transparency, Rio de Jannero, Brazil, 2024) 2; Verdadero and others (n 63) 12.

[65] Sasha Luccioni, Bruna Trevelin and Margaret Mitchell, 'The Environmental Impacts of AI -- Primer' (*Huggin Face*, 3 September 2024) <https://huggingface.co/blog/sasha/ai-environment-primer> accessed 8 June 2025.

[66] Example refers to GPT-3, Pengfei Li, Jianyi Yang, Mohammad A Islam, Shaolei Ren, 'Making AI Less "Thirsty": Uncovering and Addressing the Secret Water Footprint of AI Models' (Communications of the ACM, 17 June 2025) 2 <https://cacm.acm.org/sustainability-and-computing/making-ai-less-thirsty/> accessed 23 June 2025.

[67] Felicity Beringer, 'Thirsty for power and water, AI-crunching data centers sprout across the West' (*Stanford University Bill Lane Center for the American West*, 8 April 2025) < https://andthewest.stanford.edu/2025/thirsty-for-power-and-water-ai-crunching-data-centers-sprout-across-the-west/> accessed 24 April 2026.

[68] David Mytton, 'Data Centre Water Consumption' (2021) 4 (11) npj Clean Water 1.

[69] Miguel Yañez-Barnuevo, 'Data Centers and Water Consumption' (*Environmental and Energy Study Institute (EESI)*, 25 June 2025) <https://www.eesi.org/articles/view/data-centers-and-water-consumption> accessed 22 October 2025.

[70] Rich Kenny, 'Report: Water Use in AI and Data Centres' (Government Digital Sustainability Alliance (GDSA), 2025) <https://sustainableict.blog.gov.uk/2025/09/17/ais-thirst-for-water/> accessed 19 January 2026;

## 2.2. The efficiency paradox in hyperscale data centres

Two observations indicate why PUE and WUE scores do not reflect actual data centre efficiency. The first issue is that any initial savings in lowering PUE and WUE scores due to technical optimisations, produce diminishing returns, thereby producing a smaller (i.e., marginal) gains in efficiency over time. The second issue is that for the data centre operator will *need to* expand more high-density servers and advanced cooling technologies to sustain the increased computing demands. To sustain these claims, we note the interplay between (i) the aggregate effects of technical optimisations yielding diminishing returns of data centre efficiency, and (ii) the ratio-effects in PUE and WUE scoring and how reporting of individual data centre efficiency will only promote marginal efficiency gains.

### 2.2.1. On diminishing returns

Technical optimisations of IT equipment and infrastructure are only efficient if their gains can be sustained over a period of time. Virtualisation increases the utilisation rate of servers doing 'useful work', such as relocating workloads or reducing the power consumption of server doing background work (i.e., idle power consumption),[74] as well as reducing the number of servers in a data centre, thereby increasing energy efficiency.[75] Another approach to improve energy efficiency entails replacing conventional servers utilising Central Processing Units (CPUs) with AI-specific accelerators including GPUs and Tensor Processing Units (TPUs) to speed up model training.[76] AI-accelerators can also speed up AI inference by grouping several requests to a generative AI model in a "batch" and

---

OECD.AI Expert Group on AI Compute and Climate, 'Measuring the Environmental Impacts of Artificial Intelligence Compute and Applications' (OECD, 2022) 5 < https://www.oecd.org/en/publications/measuring-the-environmental-impacts-of-artificial-intelligence-compute-and-applications_7babf571-en.html> accessed 26 May 2025.

[71] David Mytton and Masaō Ashtine, 'Sources of Data Center Energy Estimates: A Comprehensive Review' (2022) 6 (9) Joule 2032, 2032; OECD.AI Expert Group on AI Compute and Climate (n 70) 26; Cf David Mytton, Masaō Ashtine, Scot Wheeler, David Wallom, 'Stretched Grid? Managing Data Center Energy Demand and Grid Capacity' [2023] 2 Oxford Open Energy 1, 2.

[72] Raluca Besliu, Aniket Narawad and Anna Toniolo, 'Infrastructure or Intrusion? Europe's Conflicted Data Center Expansion' (*AlgorithmWatch*, 25 July 2025) <https://algorithmwatch.org/en/infrastructure-intrusion-conflict-data-center/> accessed 19 January 2026.

[73] E.g., 'Measuring Datacenter Energy and Water Use to Improve Microsoft Cloud Sustainability' (*Microsoft Datacenters*) <https://datacenters.microsoft.com/sustainability/efficiency/> accessed 19 January 2026.

[74] Coroamă and Hinterholzer (n 24) s 7.2.3; Lynn H Kaack, Priya L. Donti, Emma Strubell, George Kamiya, Felix Creutzig and David Rolnick, 'Aligning Artificial Intelligence with Climate Change Mitigation' [2022] 12 Nature Climate Change 518, 520–521; Josh Whitney and Pierre Delfroge, 'Data Centre Efficiency Assessment' (National Resources Defence Council, 2014) 14 < https://www.nrdc.org/sites/default/files/data-center-efficiency-assessment-IP.pdf > accessed 20 April 2026.

[75] Ginelle Greene-Dewasmes, Michael Higgins, Thapelo Tladi, 'Artificial Intelligence's Energy Paradox: Balancing Challenges and Opportunities' (World Economic Forum and Accenture 2025) 5, 10 <https://reports.weforum.org/docs/WEF_Artificial_Intelligences_Energy_Paradox_2025.pdf> accessed 5 November 2025.

[76] Le Goff and others (n 9) 20; R Colin Johnson, 'AI's Increasing Power Needs' (*Communications of the ACM*, 14 August 2024) <https://cacm.acm.org/news/ais-increasing-power-needs/> accessed 24 October 2025.

processing them simultaneously for more efficient use of hardware resources, and less energy per inference task.[77]

However, aggregate effects of technical optimisations suffer from a decreasing rate of efficiency gains due to rising computing demands.[78] Roucy-Rochegoude and Buffard notes a plateauing of PUE improvements since 2023 despite technical improvements in IT equipment and cooling.[79] Reports by Masanet *et al* and Le Goff *et al* underline that improvements, such as in virtualisation and AI-accelerators, are reaching their limits, with their efficiency levelling off.[80] Further, D'Ambrosio *et al* estimate that batching may induce diminishing returns in energy efficiency when the batch size increases the energy per prompt.[81] Following this narrative, initial gains in PUE aggregate scores are counterbalanced by an increase in the overall electricity demand of AI and generative AI models.

A similar trend can be observed with regard to data centre infrastructure, as well as trade-offs between PUE and WUE yielding smaller efficiency gains. Indeed, as noted by Coroamă and Dumbravă the resulting PUE efficiencies in infrastructure 'has already achieved a very high level' with 'future gains likely only [being] marginal'.[82] Moreover, traditional air-cooling methods are typically not sufficient too cool AI servers effectively.[83] Trade-off improvements in PUE and WUE become smaller as a result due to the reliance on water-intensive methods and slower gains in efficiency.[84]

Data centre operators can only maintain marginal gains in PUE and WUE scores by further optimising for technical efficiency, which in turn, requires further retrofitting and expanding of its facility. Balancing between energy and service demand is currently matched by improvements in cooling infrastructure.[85] In this respect, Hinterholzer, Hintemann, Beucker *et al*'s technical report notes a weighted WUE average that possibly is due to 'a shift of larger facilities with higher water needs', while a lower PUE average in larger facilities reflect 'economies of scale', among others.[86] These considerations further underscore the metrics' ratio-effects, creating an efficiency paradox that spurs further growth in the hyperscale data centre sector and environmental degradation, while enabling low PUE and WUE reporting.

---

[77] Ishnet Kaur Dua, 'Model Training and Inference' in Ishneet Kaur Dua and Parth Girish Patel (eds), *Optimizing Generative AI Workloads for Sustainability: Balancing Performance and Environmental Impact in Generative AI* (APress 2024) 185-186.

[78] Commins and Irion (n 1) 25; COMMISSION STAFF WORKING DOCUMENT Implementation of the Digital Decade objectives and the Digital Rights and Principles Accompanying the document Communication from the Commission to the European Parliament, the Council, the European Economic and Social Committee and the Committee of the Regions Report on the state of the Digital Decade 2023 [2023] SWD/2023/570 final, Section 2.3.1; cf Lynn Kaack and others (n 74) 20.

[79] De Roucy-Rochegonde and Buffard (n 58) 11–12.

[80] Eric Masanet, Arman Shehabi, Nuoa Lei, Sarah Smith and Jonathan Koomey, 'Recalibrating Global Data Center Energy-Use Estimates' (2020) 367(6481) Science 984, 984; Le Goff and others (n 9) 23–24; Coroamă and Hinterholzer (n 24) 50–51; Cf Tanqi Xiao, Francesco Fuso Nerini, H Damon Matthews, Massimo Tavoni and Fengqi You, 'Environmental impact and net-zero pathways for sustainable artificial intelligence servers in the USA' [2025] 8 Nature Sustainability 1541, 1542.

[81] D'Ambrosio and others (n 4) 47; Jared Fernandez, Clara Na, Vashisth Tiwari, Yonatan Bisk, Sasha Luccioni, Emma Strubell, 'Energy Considerations of Large Language Model Inference and Efficiency Optimizations' in Wanxiang Che, Joyce Nabende, Ekaterina Shutova, Mohammad Taher Pilehvar (eds), *Proceedings of the 63rd Annual Meeting of the Association for Computational Linguistics* (Association for Computational Linguistics 2025) 9.

[82] Vlad C. Coroamă and Oana Dumbravă, 'Circular Economy Synergies and Trade-offs in Data Centres' (CEST-DC, 2026) 17 < https://arxiv.org/abs/2608.04571> accessed 2 October 2026.

[83] Kenny (n 67) 16; Coroamă and Hinterholzer (n 24) 51.

[84] Gamazaychikov and Luccioni (n 57).

[85] Manuel Herrera, Xiang Xie, Andrea Menapace, Ariele Zanfei and Bruno M Brentan, 'Sustainale AI infrastructure: A scenario-based forecast of water footprint under uncertainty' [2025] 526 Journal of Clearner Production 1, 4. In this respect, see also, George Kamiya and Oskar Kvarnström, 'Data centres and energy- from global headlines to local headaches?' (*IEA*, 20 December 2019) < www.iea.org/commentaries/data-centres-and-energy-from-global-headlines-to-local-headaches> accessed 2 October 2026.

Rebounds may further reinforce diminishing returns.[87] A rebound effect occurs when the behaviour of the operator, consumer or user changes in a way that undermines initial efficiency gains.[88] Luccioni, Strubell and Crawford note how improvements in AI hardware or inference efficiency, may stimulate additional demand for AI capabilities, thereby undermining the initial efficiency gains.[89] While neither PUE nor WUE can capture these indirect effects, noting these dynamic changes in AI usage and demand is relevant to assessing the technology's overall environmental impact, particularly when these second-order effects scale faster than data centre efficiency gains.[90] As rightly pointed out by Aczel *et al*, 'efficiency gains *alone*' do not shed light on the '*absolute reductions* in energy use' (emphasis added) and indirect effects and demand changes are important to contextualise efficiency improvements.[91]

### 2.2.2. On the efficiency paradox behind the PUE and WUE scores

To elaborate on the efficiency paradox, imagine a scenario where a data centre operator needs to match efficiency improvements to rising computational demand. The organisation might apply virtualisation to reduce the number of servers, *but will have to be scaling up* in AI-accelerators per server to handle generative AI workloads.[92] Similarly, batching of *large* input-output sequences will increase power densities.[93] Because both the PUE and WUE are ratio-metrics, the reduction in the IT load needs to be matched with an aggressive optimisation of the infrastructure to limits overhead in PUE.[94] To further optimise PUE and WUE, hyperscale operators may match their increased AI workload with liquid cooling methods.[95] This requires significant upfront costs to ensure the IT equipment's compatibility with liquid coolants dissipating the heat from the server and/or GPUs.[96]

[86] Hinterholzer and others (n 19) 41, 45; Cf European Data Centre Association, 'State of European Data Centres' (EUCDA, 2026) 43 <https://www.eudca.org/documents/content/ZlZXb4bRSRefaEVqya2I_Aoea?download=0> accessed 4 October 2026.

[87] Le Goff and others (n 9) 21, 65; OECD.AI Expert Group on AI Compute and Climate (n 70) 32, Commins and Irion (n 1) 24-25. See also, Gael Varoquax, Sasha Luccioni, Meredith Whittaker, 'Hype, Sustainability, and the Price of the Bigger-is-Better Paradigm in AI' (Proceedings of the 2025 ACM Conference on Fairness, Accountability, and Transparency (Association for Computing Machinery, Athens, Greece, 2025) 65 <https://dl.acm.org/doi/10.1145/3715275.3732006> accessed 20 April 2026.

[88] Gailhofer and others (n 21) 10.

[89] Luccioni, Strubell and Crawford (n 10) 1,5.

[90] D'Ambrosio and others (n 4) 70; Le Goff and others (n 9) 10; Luccioni, Strubell and Crawford (n 10) 1.

[91] Miriam Aczel, Sanaz Chamanara, Mir Matin, Aria Farsi, Tshilidzi Marwala and Kaveh Madani, 'Environmental Cost of AI's Energy Use: Carbon, Water and Land Footprints' (United Nations University Institute for Water, Environment and Health, 2026) 30 < https://unu.edu/inweh/collection/environmental-cost-of-AIs-Enrgy-Use-Carbon-water-and-land-footprints> accessed 2 October 2026.

[92] D'Ambrosio and others (n 4) 62; Beth Kindig, 'AI Power Consumption: Rapidly Becoming Mission-Critical' (*Forbes*, 20 June 2024) <https://www.forbes.com/sites/bethkindig/2024/06/20/ai-power-consumption-rapidly-becoming-mission-critical/> accessed 24 October 2025.

[93] Mauricio Fadel Argerich and Marta Patiño-Martínez, 'Measuring and Improving the Energy Efficiency of Large Language Models Inference' [2024] 12 IEEE Access 80194, 80202; Francisco Caravaca, 'Measuring Energy Consumption of LLMs Inferences' (2026) 53 (3) ACM SIGMETRICS Performance Evaluation Review 18; Jesse Dodge, Taylor Prewitt, Remi Tachet des Combes, Erika Odmark, Roy Schwartz, Emma Strubell, Alexandra Sasha Luccioni, Noah A Smith, Nicole De Cario and Will Buchanan, 'Measuring the Carbon Intensity of AI in Cloud Instances' (Proceedings of the 2022 ACM Conference on Fairness, Accountability, and Transparency (Association for Computing Machinery, Seoul, Korea, 2022) 1879–1880.

[94] BS EN 50600-4-2:2016+A1:2019 (n 8) s D.2; Brady and others (n 40) 160–161; see also, Md Abu Bakar Siddik, Arman Shehabi and Landon Marston, 'The environmental footprint of data centers in the United States' (2021) 16 (6) Environmental Research Letters 1, 3.

These current considerations, while not raising PUE and WUE, do not improve the status quo of minimising the energy footprint of hyperscale data centres either. Rather, these maintain a small, marginal gain in efficiency by increasing both IT equipment and infrastructure. This way, hyperscale data centres may be able to still report "efficient" PUE and WUE scores, while masking that this amount of scaling comes with an increasing ecological price tag on electricity supply and water stress. In this regard, Freitag *et al* further note that aggressive processor-level efficiency improvements might come to an end, effectively leaving no counterbalance to the rapid demand and energy use.[97] Hence, we see that in addition to the discussed PUE/ WUE operational scope, their complex interdependencies producing aggregate effects of technical optimisation, as well as the metrics' ratio-effects effectively serve Big Tech's market-incentives. These incentives entail keeping up with new requirements in IT load, power densities and cooling in large generative AI workloads, conditioning the further expansion of hyperscale data centres.

# 3. A new data centre efficiency strategy and the recast EU Energy Efficiency Directive

Addressing the efficiency paradox in PUE and WUE reporting requires a new data efficiency strategy. This new strategy should fulfil two aims: (i) for data centre operators to demonstrate the actual ecological benefits of efficiency improvements, and (ii) for other regulators to have the necessary context to intervene when efficiency improvements yield diminishing returns. Both aims are important in light of policy interventions within Article 12 of the recast EED including its reporting framework, labelling, future checks and balances and thresholds.[98]

An ecological benefit requires from data centre operators and regulators including the EU Commission forward-looking planning to put 'increased' efficiency gains into context.[99] An efficiency gain is demonstrated on two levels:

- **(I)** A PUE and WUE reporting framework where data centre operators demonstrate actual efficiency gains;
- **(II)** Policy interventions targeted to outpace the diminishing returns of aggregate PUE and WUE efficiency improvements.

---

[95] Amy S Fleischer, 'Cooling our insatiable demand for data' (2020) 370 (6518) Science 783- 784; Patel and others (n 59) 13; Shehabi and others (n 48) 44; Gröger and others (n 44) 23; Carolina Koronen, Max Åhman and Lars Nilsson, 'Data centres in future European energy systems – energy efficiency, integration and policy' [2020] Energy Efficiency 129, 133.

[96] United Nations Environment Programme (UNEP) United for Efficiency (U4E) initiative (n 27) 18, 30; Yañez-Barnuevo (n 69).

[97] Charlotte Freitag, Mike Berners-Lee, Kelly Widdicks, Bran Knowles, Gordon S Blair and Adrian Friday, 'The real climate and transformative impact of ICT: A critique of estimates, trends, and regulations' (2021) 2 (1) Patterns 1, 5.

[98] *recast EED* (n 7) art 12; Commission Delegated Regulation of 21.9.2026 supplementing Directive (EU) 2023/1791 of the European Parliament and of the Council and amending Commission Delegated Regulation (EU) 2024/1364 as regards the establishment of a common Union rating scheme for data centres [2026] C(2026) 3472 final (hereafter Proposed Commission Delegated Regulation 2026); Commission Delegated Regulation (EU) 2024/1364 of 14 March 2024 on the first phase of the establishment of a common Union rating scheme for data centres [2024] OJ 2024/1364 (hereafter Commission Delegated Regulation 2024/1364 of 14 March 2024).

[99] Focus Group on Environmental Efficiency for Artificial Intelligence and other Emerging Technologies (FG-AI4EE), 'Requirements on Energy Efficiency Measurement Models and the Role of AI and Big Data' (Working Group 2) (2021) (D.WG2-03) at 3.1.15 < https://www.itu.int/en/ITU-T/focusgroups/ai4ee/Documents/T-FG-AI4EE-2021-D.WG2.03-PDF-E.pdf > accessed 22 January 2026.

This strategy on ecological benefits maintains the interplay between individual, ratio-based PUE and WUE consequences, as well as systemic effects outpacing efficiency gains. Data centre operators need to provide additional details on how positive efficiency improvements produce water and energy trade-off improvements when increasing IT load and cooling requirements. These contextual details elaborate on the PUE and WUE ratio and how efficiency improvements address the status quo of moving beyond marginal gains in efficiency improvements. Positive efficiency improvements also need to be reflected in the context around water use, focusing on water stress including details about water source (i.e., the use of potable or processed water), as well as water accessibility.[100] The data centre operator would need to elaborate on their efficiency improvements in cooling technology and why less water-intensive methods were (not) adopted alongside the PUE score.

Reported PUE and WUE values will serve the EU Commission to formulate policy interventions when ecological gains are levelling off, considering indirect effects including diminishing returns. The visibility of the ecological gains in PUE and WUE reporting is important for policy interventions. In this respect, documenting ecological gains can help regulators and potentially other stakeholders (e.g., intergovernmental, non-profit, and research organisations)[101] who gain access to the data to examine the turning point of diminishing returns. For example, empirical data on how hyperscale data centres select technical optimisation and cooling technologies can help stakeholders to investigate rebound effects and environmental impact assessments, [102] and can likewise offer an evidence-base in policy for incentivising new technical optimisation methods and infrastructure, such as encouraging more energy-efficient on-device processors.[103]

## 3.1. Ecological gains in the recast EU Energy Efficiency Directive

Both aspects of our strategy, which require the documentation of ecological gains and establish intervention points based on diminishing returns, can be incorporated into existing EU policy on data centre efficiency. This alignment is particularly significant given that the EU is supporting substantial hyperscale data centre growth while simultaneously aiming for energy-efficiency improvements.[104]

The recast EED is best understood as a regulatory instrument designed to create a baseline for future regulatory interventions and a common rating scheme on the basis of reporting of key metrics.[105] Article 12 stipulates a mandatory reporting framework for data centre operators of facilities with a 500kw+ power demand[106] to report key metrics including aggregated PUE and WUE values to a European database.[107] The first Delegated Regulation of the recast EED (hereafter: *Delegated Regulation 2024*) further underlines that measurement of PUE and WUE values shall adopt the methodology of relevant CEN/CENELEC standards.[108] Two technical reports are building on the first reporting period in 2024, and make recommendations for additional policy interventions, such as harmonisation of reporting standards and further clarification of reporting requirements.[109] Further, the European Code of Conduct on Data Centre Energy Efficiency (the EU CoC) recommends data centre operators to adopt efficiency improvements and an energy management plan, alongside reporting PUE and WUE values.[110] Signatories, which include some data centres operated by Facebook Ireland LTD, Microsoft Corporation and Google Data Centres in the EU, may use the EU CoC as an assessment tool to report on the PUE and WUE values.[111] The Taxonomy Climate Delegated Act provides a classification scheme for data centres' economic activities which can be independently audited for compliance with the EU CoC.[112]

---

[100] Alex Setmajer, ‘How Data Centers Use Water, and How We’re Working to Use Water Responsibly’ (*Interconnections - The Equinix Blog*, 19 September 2024) <https://blog.equinix.com/blog/2024/09/19/how-data-centers-use-water-and-how-were-working-to-use-water-responsibly/> accessed 10 November 2025. Again, see also Hacker (n 6) 354-356; Hacker and others ‘Policy Brief on Powering Europe’s Digital Transformation: A Roadmap to Greener Centers’ (n 6) 3.

[101] Cf Thomas Le Goff, ‘Not Greenwashing, but Still… A Closer Look at Big Tech’s 2025 Sustainability Reports’ [2025] Internet Policy Review 1.

[102] In this regard, see German Federal Government, ‘Artificial Intelligence Strategy of the German Federal Government 2020 Update’ (2020) 16 <https://www.ki-strategie-deutschland.de/files/downloads/Fortschreibung_KI-Strategie_engl.pdf> accessed 22 January 2026; EU Commission, ‘Report on the Digital Decade’ (27 September 2023) 35 < https://digital-strategy.ec.europa.eu/en/library/2023-report-state-digital-decade> accessed 12 November 2025.

[103] E.g., D’Ambrosio and others (n 4) 71; Cf Peter Howson, ‘DeepSeek Claims to Have Cured AI’s Environmental Headache. The Jevons Paradox Suggests It Might Make Things Worse’ (*The Conversation*, 31 January 2025) <https://doi.org/10.64628/AB.yq6p6kgc9> accessed 11 November 2025.

[104] Communication from the Commission to the European Parliament, The Council, The European Economic and Social Committee and the Committee of the Regions: AI Continent Action Plan [2025] COM/2025/165 final, s. 1.3; *recast EED* (n 7) Recital 2.

[105] Viegand Maagoe and COWI (n 53) 8–9; Ebert and others ‘AI, Climate, and Regulation: From Data Centers to the AI Act’ (n 6) 3.

Building on the reporting scheme, the EU Commission, while underscoring the inconsistent disclosure and high variability in PUE and WUE scores,[113] has developed a labelling and classification framework.[114] The most recent proposed Commission Delegated Regulation (hereafter: *proposed Delegated Regulation 2026*) envisages a label that would display PUE and WUE's contextual values, as well as their efficiency classes.[115] In addition, the EU Commission will prepare a document that briefly contextualises PUE and WUE readings, with the latter's relative importance on water stress.[116] Finally, it amends the WUE's terminology, replacing the term of 'potable water input' with the notion of ‘freshwater input’ considering Annex II of the Water Framework Directive.[117] Altogether, this baseline on transparency, context and harmonisation has been set out to inform its future-oriented objectives for minimum performance standards, and reviews in the Cloud and AI Development Act, and the Climate and Environmental Delegated Acts under the Taxonomy Regulation.[118]

---

[106] On criticism of the 500kw benchmark, see Hacker and others 'Policy Brief on Powering Europe's Digital Transformation: A Roadmap to Greener Data Centers' (n 6) 3.
[107] *recast EED* (n 7) arts 12; Annex VII.
[108] *Commission Delegated Regulation 2024/1364 of 14 March 2024* (n 98) Annex II, Annex III (a)-(b).
[109] Heatubun and others (n 14) 54-55; Hinterholzer and others (n 19) 68-69.
[110] Heatubun and others (n 14) 19; Brocklehurst (n 14) 68.
[111] Mark Acton, John Booth and Daniele Paci, '2025 Best Practice Guidelines for the EU Code of Conduct on Data Centre Energy Efficiency' (*JRC Publications Repository*, 2025) 3 <https://doi.org/10.2760/9449356> accessed 1 September 2025.
[112] Commission Delegated Regulation (EU) 2021/2139 of 4 June 2021 supplementing Regulation (EU) 2020/852 of the European Parliament and of the Council by establishing the technical screening criteria for determining the conditions under which an economic activity qualifies as contributing substantially to climate change mitigation or climate change adaptation and for determining whether that economic activity causes no significant harm to any of the other environmental objectives [2021] OJ L 442/1 (hereafter Commission Delegated Regulation (EU) 2021/2139) Annex I, Section 8.1; EU Commission, 'Commission Notice on the interpretation and implementation of certain legal provisions of the EU Taxonomy Climate Delegated Act establishing technical screening criteria for economic activities that contribute substantially to climate change mitigation or climate change adaptation and do no significant harm to other environmental objective' [2023] C/2023/6756, para 160. Cf Annex I to the Commission Delegated Regulation amending Delegated Regulation (EU) 2021/2139 as regards enhancing the usability of the technical screening criteria [2026] Ares(2026)2879622 (hereafter Annex I to the Commission Delegated Regulation amending Delegated Regulation (EU) 2021/2139), page 145, Section 1 (c).
[113] EU Commission, 'Report from the Commission to the European Parliament and the Council on the energy efficiency of data centres in the EU' [2026] COM(2026) 500 final, pages 4- 8.
[114] ibid pages 14-15.
[115] *Proposed Commission Delegated Regulation 2026* (n 98) art 3.
[116] Annexes to the Commission Delegated Regulation (EU) supplementing Directive (EU) 2023/1791 of the European Parliament and of the Council and amending Commission Delegated Regulation (EU) 2024/1364 as regards the establishment of a common Union rating scheme for data centres [2026] C(2026) 3472 final (hereafter Annexes to the Commission Delegated Regulation (EU) supplementing Directive (EU) 2023/1791), Annex II (2)- (3).

While Article 12 of the recast EED require data centre operators to disclose their PUE and WUE scores, they are not obliged to contextualise these efficiency improvements within the metric’s trade-offs. Moreover, the EU Commission does not clarify how it would ensure consistent PUE and WUE reporting. This is important as these scores require contextual information for policy interventions to incentivise ecological benefits over marginal scoring gains. Finally, the EU Commission does not have a mechanism in place for tracking diminishing returns and rebound effects.

### 3.1.1 Proposals for improving the recast EU Energy Efficiency Framework

Focusing on limitations in the current recast EED framework, we suggest three concrete proposals incorporating the main elements of our new data centre efficiency strategy. First, the EU Commission should adopt a Delegated Regulation on efficiency improvements to clarify PUE/WUE interdependencies and trade-offs. Second, we suggest that minimum performance standards provide a good stepping stone for ensuring evidence-based comparisons in reporting and incorporating water stress. Third, the EU Commission should track efficiency improvements over time, considering their investments in the data centre sector and large-scale facilities for the training and deployment of Large Generative Models.

#### *3.1.1.(a) Proposal I: on adopting a Delegated Regulation on efficiency improvements*

Articles 12 and 33 (3) underscore the EU Commission’s incremental approach to establish a common rating scheme for data centres, entailing the framework on the reporting of PUE/ WUE values, as well as their documentation and rating in the electronic label.[119] In this respect, the proposed Delegated Regulation 2026 sets out the metrics’ interdependencies and leaves open contextualising PUE/WUE values, taking into account their ‘trade-offs’ and ‘possible ways to rate sustainability’ in Annex III (3) (III) and (VII).[120] To fill this gap, we propose for the EU Commission to adopt a Delegated Regulation to ensure enhanced documentation of PUE and WUE’s trade-offs relative to efficiency improvements.

The current system is only a first step towards meaningful transparency of PUE and WUE trade-offs including their limitations. Annex III (3) notes that ‘[e]ach label shall be accompanied by a document common to all labels, that will be prepared by the EU Commission’ and ‘*shortly* explain’ a set of elements (emphasis added).[121] One element includes the PUE/ WUE interdependencies with the facility’s size, and the PUE’s relationship to climate conditions, geography, their seasonal variability.[122] However, this element includes PUE/WUE interdependencies *and* trade-offs.[123] Hence, the EU Commission, when preparing the accompanying document to the label, needs to ensure that trade-offs are documented separately, not deduced from comparing PUE and WUE’s relative values. This requires enhanced documentation of PUE and WUE trade-offs, which need to be relative to the efficiency gains in choosing and alternating between different cooling technologies. Additionally, documenting the PUE and WUE ratio effects requires enhanced disclosure on any infrastructure upgrades and technical optimisations in overall IT load,[124] and guidance on new indicators.[125] The additional context would need to also inform the review of the Taxonomy Climate Delegated Act, informing third-party auditing of the 'expected practices' within the EU CoC *in addition to* the energy efficiency classes' scheme.[126] Disclosing the ratio-effects in PUE/WUE trade-offs will avoid distortions to other elements in the accompanying document, including 'possible ways to rate sustainability' in Annex II (3) (VII).**[127]** An additional Delegated Regulation would need clarify both aspects on deducing the relative nature of PUE/WUE's, revealing whether their trade-off is appropriate, thereby moving beyond their mere interdependencies, and considering their limitations and scope of efficiency improvements.

---

[117] *Proposed Commission Delegated Regulation 2026* (n 98) art.5 (17); Directive 2000/60/EC of the European Parliament and of the Council of 23 October 2000 establishing a framework for Community action in the field of water policy [2000] OJ L 327/1 (Water Framework Directive), Annex II, 1.2.3 System A. 1.2.4 System A.
[118] *Proposed Commission Delegated Regulation 2026* (n 98).
[119] *recast EED* (n 7) arts 12 and 33.
[120] *Annexes to the Commission Delegated Regulation (EU) supplementing Directive (EU) 2023/1791* (n 116) Annex (3) (III) (VII).
[121] ibid Annex (3) (III).
[122] *Annexes to the Commission Delegated Regulation (EU) supplementing Directive (EU) 2023/1791* (n 116) Annex II (2) (XIV), Annex II (3) (I) -(VI)
[123] ibid Annex II (3) (III).

#### *3.1.1.(b) Proposal II: on clarifying the role of minimum performance standards*

Uniform reporting of PUE and WUE values is not only important to avoid reporting gaps,[128] but also secure evidence-based comparisons. Minimum performance standards would need to go beyond mere PUE/WUE thresholds, and should set a comparative baseline for efficiency improvements securing actual ecological gains, considering water stress**.**

The 2024 and 2026 reports**,** while identifying gaps in PUE and WUE reporting fails to provide a forward-looking approach that secures consistent and transparent reporting. The 2024 technical report notes a 'high variability' in reported PUE values across Member States, with a significant trend that smaller data centres generally show higher values, while larger centres report lower scores.[129] The report further highlights large differences in WUE values, a lack of contextual information surrounding the true water impact, and trade-offs in PUE and WUE scores.[130] The second technical report then suggests targets for the PUE of 1.3 for new data centres and < 0.4 $m^3$/MWh for the WUE.[131] However, the 2026 report then notes that motes data centres did not attain this WUE target and exceed a recommended PUE baselines.[132] Yet, the proposed system establishes energy efficiency classes incorporating the aforementioned and further value ranges.[133] According to the EU Commission, minimum performance standards could envisage PUE/WUE thresholds to further 'encourage industry-wide advancements'**.** While PUE/WUE thresholds and targets might be useful to 'remove the worst-performing facilities from the market'[134] by suggesting common energy efficiency endpoints,[135] they do not provide a remedy for consistent highlighting of marginal efficiency gains. Hence, policy interventions based on these energy classes and future minimum performance standards will be inconsistent and ineffective indicators where operators demonstrate the ecological gains of data centre efficiency.

Future guidance needs to set out a framework that aims for consistent and comprehensive methodologies for data centre operators to showcase ecological gains for efficiency improvements. Focusing on the WUE, this requires more granular documentation beyond input, focusing on the relative importance 'to the levels of water stress'. [136] At the moment, the EU Commission still refers to the CEN/CENELEC standard that includes different approaches to calculate the WUE metric, with leeway to operators to include or exclude potable and non-potable sources.[137] The proposed Delegated Regulation 2026 expands to non-portal water sources (i.e., freshwater) based on a salinity level of <0.5 within the Water Framework Directive**.**[138] However, what is missing is a clear terminology for data centre operators to provide a breakdown of all water sources and the outline of different WUE values with seasonal water demand. This lack of context undermines evidence-based comparison across different WUE values as a basis for stipulating energy classes and future minimum performance standards. Hence, the EU Commission needs to formulate additional guidance on how operators should disclose contextual information on documenting water stress levels. This can include a breakdown of the water sources and the ecological gain in relying on different strategies, such as reducing pressure on potable supplies or improving local resilience.

[124] In this respect, see *Proposed Commission Delegated Regulation 2026* (n 98) Annex III (3) (e).
[125] cf ibid art 6 (d).
[126] Cf *Annex I to* the *Commission Delegated Regulation amending Delegated Regulation (EU) 2021/2139* (n 112) page 145, Section 1 (b) and (c); *Commission Delegated Regulation* (EU) 2021/2139 (n 112) s Annex I, Section 8.1 (1); cf Acton, Booth and Paci (n 111), s.2.5.
[127] *Annexes to the Commission Delegated Regulation (EU) supplementing Directive (EU) 2023/1791* (n 116) Annex II (3) (VII).
[128] Hinterholzer and others (n 19) 23.
[129] Heatubun and others (n 14) 34.
[130] Heatubun and others (n 14) 40-41; Hinterholzer and others (n 19) 42.
[131] Hinterholzer and others (n 19) 50.
[132] EU Commission, 'Report from the Commission to the European Parliament and the Council on the energy efficiency of data centres in the EU' (n 113) pages 7-8.
[133] *Annexes to the Commission Delegated Regulation (EU) supplementing Directive (EU) 2023/1791* (n 116) Annex I, Table 1 and Table 2.
[134] Hinterholzer and others (n 19) 67.
[135] Cf Ruschemeier (n 6) 20.

### *3.1.1. (c) Proposal III: on tracking efficiency improvements and formulating strategic priorities*

While the previous proposals focus on positive efficiency improvements, this recommendation considers the endpoints of efficiency improvements including their countereffects. This is important to balance efficiency improvements with their diminishing returns rather than market incentives. A mechanism for monitoring diminishing returns and rebounds of technical optimisation can act as evidence-based details to actively review the recast EED in light of countereffects and rebound effects.

As a first step, this requires a clearly set out mechanism for the EU Commission to include periodic assessments of efficiency improvements within Article 6 of the proposed Delegated Regulation 2026 including third-party access to reported data.[139] In this regard, a heat map can be useful for the EU Commission to isolate efficiency improvements, PUE and WUE scores across hyperscale data centres over time, as well as support spotting diminishing returns in efficiency improvements.[140] Tracking endpoints of efficiency improvements will, in turn, need to act as a baseline when reviewing the label in light of the proposed Cloud and AI Development Act's strategic priorities' to secure energy-efficient data centre scaling.[141]

Finally, balancing between the scaling of hyperscale infrastructures and energy efficiency is not only a sustainability issue but implies a political and socio-economic dimension.[142] While embedding these intersections would go beyond the paper's scope, the arguments revealing the efficiency paradox contribute to a more nuanced policy understanding of data centre efficiency**.** For instance, the EU currently invests into AI (Giga)factories with strategic priorities on the development of large generative models and public-private partnerships, such as with universities and financial institutions.[143] However, no comprehensive criteria on securing energy-efficient infrastructure have been established beyond the usage of PUE and WUE values.[144] Tracking dynamic changes in data centre energy efficiency will be a first step to examine broader objectives in innovation and sustainability informing key dimension, such as site selection[145], investments in (alternative) business models, as well as further research revealing the environmental price tag behind strategic market incentives supporting increased scale and AI demand.

[136] *Annexes to the Commission Delegated Regulation (EU) supplementing Directive (EU) 2023/1791* (n 116) Annex III (3) (VII)
[137] ibid Annex III (5) (i); see also, Heatubun and others (n 14) 39; Hinterholzer and others (n 19) 42.
[138] *Proposed Commission Delegated Regulation 2026* (n 98) Article 5 (17); *Water Framework Directive* (n 11 7).
[139] *Proposed Commission Delegated Regulation 2026* (n 98) Article 6.
[140] Hinterholzer and others (n 19) 29.
[141] *Proposed Commission Delegated Regulation 2026* (n 98) explanatory memorandum; Proposal for a Regulation of the European Parliament and of the Council establishing a framework of measures for strengthening Europe's cloud and AI ecosystem (Cloud and AI Development Act) [2026] COM(2026) 502 final.
[142] E.g., Sandra Wachter, Brent Mittelstadt, Daria Onitiu and Netta Weinstein, 'Optimising for Happiness in EU Technology Policy to Repay Public Subsidies for AI' (SSRN pre-print, 18 June 2026) < https://papers.ssrn.com/sol3/papers.cfm?abstract_id=6939638> accessed 9 October 2026; Ana Validiva and Mohammad Anwar, 'The Technopolitics of Data Centers in the Age of AI: Concepts, Methodologies, and Challenges' in Rayvon Fouché (ed), *Oxford Research Encyclopedia of Science, Technology, and Society* (OUP 2026).

# 4. Conclusion

As the EU races to increase its data centre infrastructure, examining how the exponential growth of AI and large generative models can be made sustainable exposes a fundamental conflict: technologies that initially reduce electricity and water use ultimately require further retrofitting and upscaling to accommodate AI workloads. This paper presents a novel way to work around this efficiency paradox when hyperscale operators report energy efficiencies in their PUE and WUE scores, which the EU Commission then relies on to justify additional policy interventions.

Specifically, the paper proposes a strategy for data centre operators and regulators to demonstrate and formalise the actual, ecological gains in efficiency improvements. For data centre operators, demonstrating an ecological gain means providing additional details about how positive efficiency improvements produce water and energy trade-offs when increasing IT load and cooling requirements. For the EU Commission, formalising ecological gains is having a better grasp of when efficiency improvements are levelling off, considering indirect effects including diminishing returns.

Focusing on the reporting framework in the recast EED, we make three proposals for the EU Commission to implement our new data centre efficiency strategy to balance between its current and long-term objectives in data centre expansion while respecting its energy-efficiency target. *Proposal I* supports a new Delegated Regulation to clarify PUE and WUE trade-offs relative to efficiency improvements. *Proposal II* clarifies the role of minimum performance requirements acting as a comparative baseline for efficiency improvements securing actual ecological gains and considering water stress. *Proposal III* envisages a mechanism for the EU Commission to track efficiency improvements, their diminishing returns and countereffects over time. Implementing these measures will ensure that the recast EED common rating scheme is fit-for-purpose, balancing sustainability with AI innovation *for* local communities.

# Competing interest declaration:

No competing interests to declare.

# Acknowledgements:

The authors are indebted to Professor Philipp Hacker, Dr Jake Stone and Dr Stratis Tsirtsis for the invaluable feedback on earlier versions of the paper.

---

[143] AI Continent Action Plan (n 104) s.1.1; cf Alek Tarkowski and Felix Sieker, 'European Public AI – Policy Brief' (Bertelsmann Stiftung, 2026) 7-8 <https://www.bertelsmann-stiftung.de/en/publications/publication/did/european-public-ai-policy-brief> accessed 23 January 2026.

[144] Cloud and AI Development Act (n 141) art. 11 (1).

[145] Nicoleta Kyosovska and Andrea Renda, 'EU Plans for AI (Giga) Factories: Sanctuaries of Innovation, or Cathedrals in the Desert?' (CEPS, 3 November 2025)18 <www.ceps.eu/ceps-publications/eu-plans-for-ai-gigafactories-sanctuaries-of-innovation-or-cathedrals-in-the-desert/> accessed 10 January 2026.

## Funding:

This work has been supported by the Alexander von Humboldt Foundation in the framework of the Alexander von Humboldt Professorship (Humboldt Professor of Technology and Regulation) endowed by the Federal Ministry of Education and Research via the Hasso-Plattner-Institute.